# An Enhanced Geo-Location Technique for Social Network Communication System

**ODIKWA Henry[1], IFEANYI-REUBEN Nkechi[2], THOM-MANUEL, Osaki Miller[3]**

[1] Department of Computer Science, Wellspring University, Benin City, Edo State
[2] Department of Computer Science, Rhema University Nigeria, Aba, Abia State
[3] Information and Communication Center, Ignatius Ajuri University, Port-Harcourt, Rivers

[1]dodikwa@yahoo.com, [2]ifeanyi-reubenn@rhemauniversity.edu.ng, [3]Osakimil@yahoo.com

**ABSTRACT**

Social networks have become very popular in recent years because of the increasing large number and affordability of internet enabled gadgets such as personal computers, mobile devices and internet tablets. It has been observed that the tempo of fraud in social media these days is over alarming most especially in Nigeria. As a result of this, there is need to fortify the social network services in order to secure e-mail communication and reinforce data security. This paper advocates for an advanced and secured approach for improving communication in a social Network with the use of geo-location technique. The system was designed using an Object-Oriented software development methodology and implemented using the server-based scripting language - PHP, Cascading Style Sheets (CSS) and backend with MySQL. The proposed system will help the government and security agencies fight recent security challenges in the country.

## 1. INTRODUCTION

Social network websites have multiplied broadly all over the world and are used by different users for numerous intentions [8]. They have recently become very popular these years because of the increasing affordability and proliferation of internet enabled gadgets such as personal computers, mobile devices and other more modern hardware inventions and innovations such as internet tablets. This is exemplified by the escalating recognition of many online social networks such as Twitter, Facebook and LinkedIn. Such social networks have brought about tremendous blast of network-centric data in a wide diversity of scenarios.

Social networks may be defined in terms of either in the milieu of systems such as Facebook which in turn are explicitly designed for social communications, or in provisos of websites such as Flicker, which are designed for a different service such as content sharing, but which allow an extensive level of social interaction. [2] Defined social network "as a network of communications or associations, where the nodes comprise of actors and the edges comprise of the relationships or interactions between these actors". A generalization of the idea of social networks is that of information networks, in which the nodes could comprise either actors or entities, and the edges, denote the relationships between them. The social network has become one of the basic needs for many people because people contact one another as part of their social lives through the network and also carry out their daily business activities within the social network [10].

Social network play a vital role in the development and formation of civil society in every country [9]. A lot of people cannot discharge their duties effectively without social network. As the demand for the usage of the social network is increasing, it has been observed that the rate of fraud in it these days is over alarming most especially in Nigeria. As a result of this, there is need to fortify the social network services in order to secure the system and reinforce data security. [6] emphasized the increased interest in accurate location finding and location-based applications for indoor areas.

This paper presents an advanced and secured approach for improving communication in a social Network with the use of geo-location technique. Though many systems have evolved over time on the design of social media platforms but this work adopts geo-location techniques and improved the networks, thereby enhancing the security of users due to crime and criminality recently experienced in the society.

Geo-location is the process or technique of identifying the geographical location of a person or device by means of digital information processed through the Internet. Geo-location is used to plot courses, track elevation changes, see location history, tag images on social media, get the local weather, etc. It can also be used in our everyday lives to make our things easier; to interact with people; to solve problems; to set goals and to track anything. Geo-location is used to improve communication system like websites.

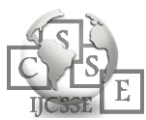

## 2. RELATED WORKS

Throughout much of human history, technologies have been developed that make it easier for people to communicate [1]. For example, the use of telegraph to transmit and receive messages over long distances dated back to 1792, [5]. Emile Durkheim, a French sociologist known by many as the father of sociology, and Ferdinand Tonnes, a German sociologist are considered pioneers of social networks during the late 1800s. Over the years, social media has evolved rapidly like LinkedIn which gained popularity in 2000. Youtube in 2005; then Facebook and Twitter by the year 2006 [4]. Many researchers have worked tirelessly to enhance social media networks by building systems that will help users enjoy the usage of social media networks.

In the design and implementation of social networking platforms for cloud deployment specialist, [7] developed a platform for cloud specialist to interact with each other. Efforts was made in enhancing social media based web application for prospective university students by creating Internet platforms for individuals, groups of people who intend gaining admission in a particular university [3]. [11] developed a face recognition online social networks as a security measure to tackle menace resulting from frauds in social media networks using face tagging done automatically. A new campus social networking systems for users was developed to enable interactions among staff and students in relation to sharing learning, working experiences and cultural life [12].

[10] suggested that social communication creates a positive impact on the social life of an individual in terms of family, education and healthcare. The result revealed that the social network has various good effects on people's social lives; people can consult professional experts such as medical practitioners, teachers and many others. Social network has formed a means of creating caring relationships among people.

[6] presented an overview of the technical aspects of the technologies for wireless indoor location systems. The major challenges for accurate location finding in indoor areas are described to be the complexity of radio propagation and the ad hoc nature of the deployed infrastructure. Fig 1 display the architectural structure of their proposed wireless geo-location systems.

In fig 1, the location metric indicates the approximate arrival direction of the signal or the approximate distance between the Mobile Terminal (MT) and Reference Point (RP); the Angle Of Arrival (AOA) – metric commonly used in direction-based systems; the Received Signal Strength (RSS); carrier signal Phase Of Arrival (POA), and Time Of Arrival (TOA) of the received signal are the metrics used for estimation of distance.

The proposed improves the functionalities of the existing systems by adding platform to curb the activities of some unscrupulous elements that lure their fellow media users into defrauding them. Creating a social media network with enhanced geo-location technique will immensely help in checkmating the activities of those who use social media platforms for negative purposes.

## 3. MATERIALS AND METHODS

The fig 2 shows the architectural design of the proposed social network system with an enhanced geo-location technique that detects In fig 1, the location metric indicates the approximate arrival direction of the signal or the approximate distance between the Mobile Terminal (MT) and Reference Point (RP); the Angle Of Arrival (AOA) – metric commonly used in direction-based systems; the Received Signal Strength (RSS); carrier signal Phase Of Arrival (POA), and Time Of Arrival (TOA) of the received signal are the current user's location in any part of the country.

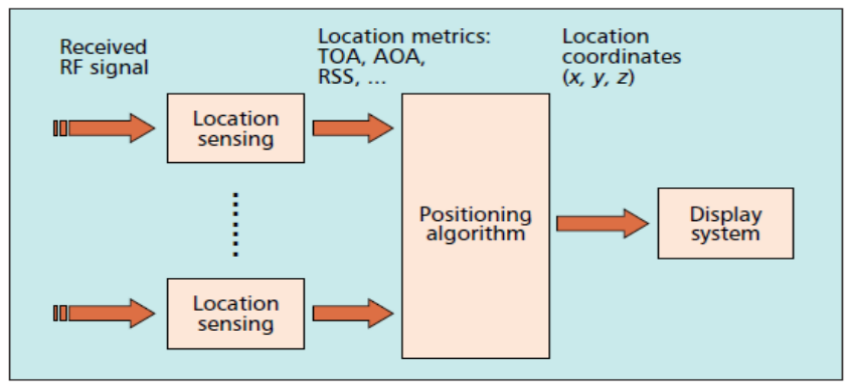

*Fig. 1. Architecture of Wireless Geo-location Systems (source: Charu, 2016)*

The proposed system is a web-based social media platform with a resilient MySQL database, enhanced security features and location maps.

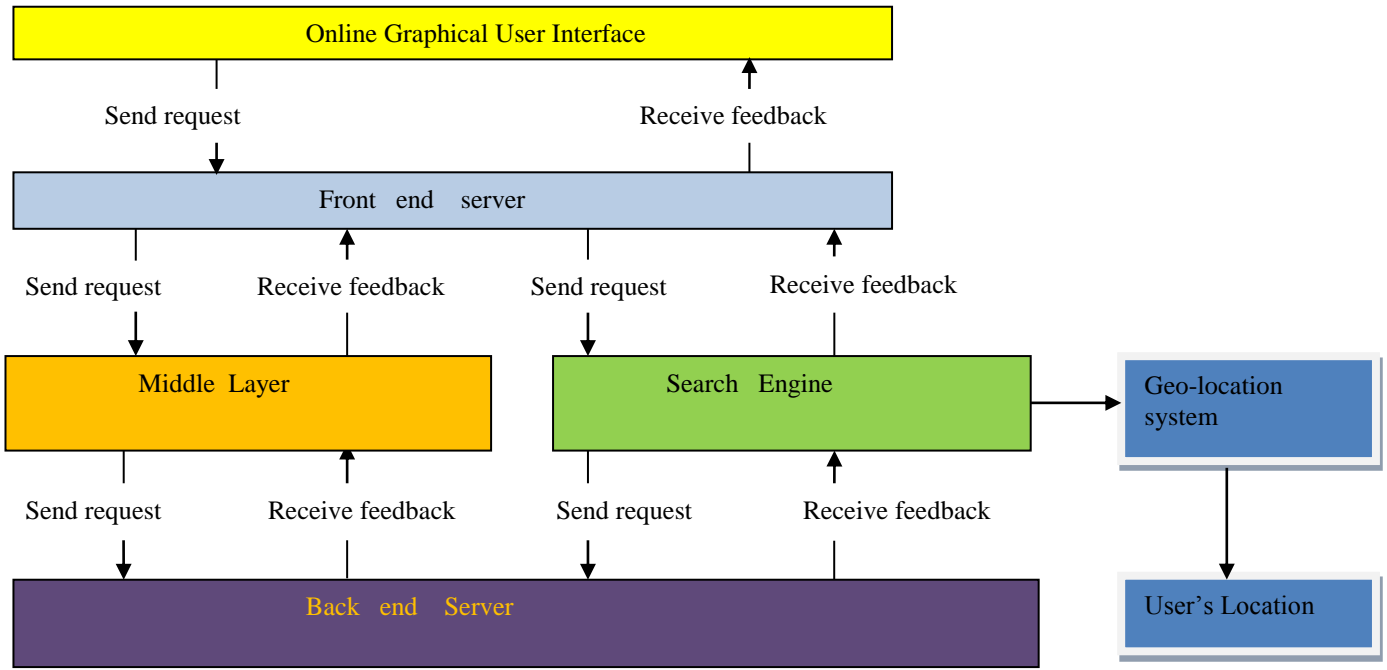

*Fig. 2. Architecture of the Proposed System*

The system is divided into three components and these are:

i. The Front End Server
ii. The Back End Server and
iii. The Geo-Location System

The functionalities are briefly described as follows:

3.1 The Front End Server

The front end server is designed to enable the clients/consumers have access to the webpage of the social media front end. This is implemented using the server-based scripting language - PHP, Cascading Style Sheets (CSS) PHP and CSS. A scripting language is a language that interprets scripts at runtime. A script is a set of programming instructions that is interpreted at runtime.

3.2. Back End Server

The back end server of the proposed system is designed with the MySQL. This is mainly for the design of the database. This includes the search engine that searches the profiles of the users on the platform and display them accordingly.

3.3 The Geo-Location System

The geo-location technique employed for the design system has the Automatic Location Identification Database (ALI) with authentic location information that exploits Google maps to display the current location of any user on the social media platform. The sequence diagram of the geo-location system is shown in fig 3. The behavioural UML is used to analyze the various components that interact in the geo-location social media network system.

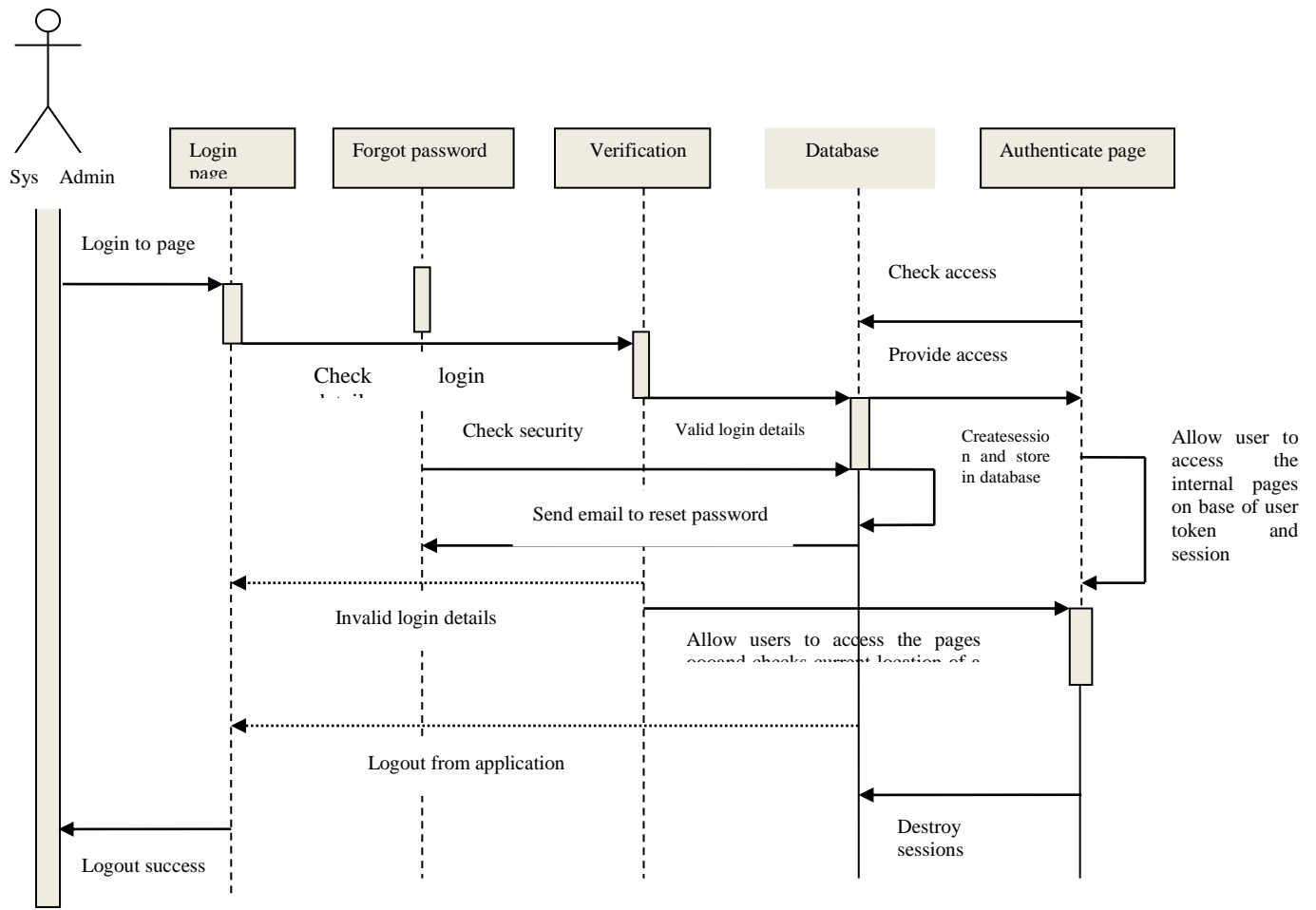

*Fig. 3. Sequence Diagram of the Proposed System.*

Fig 3 shows the sequence diagram of the proposed geo-location social network system which depicts the sequence of messages and interactions among the actors in the system.

The system administrator, the actor creates the login parameters, checks the verification and security access of the user. UML Sequence Diagrams shows the interactions and details on how operations are carried out.

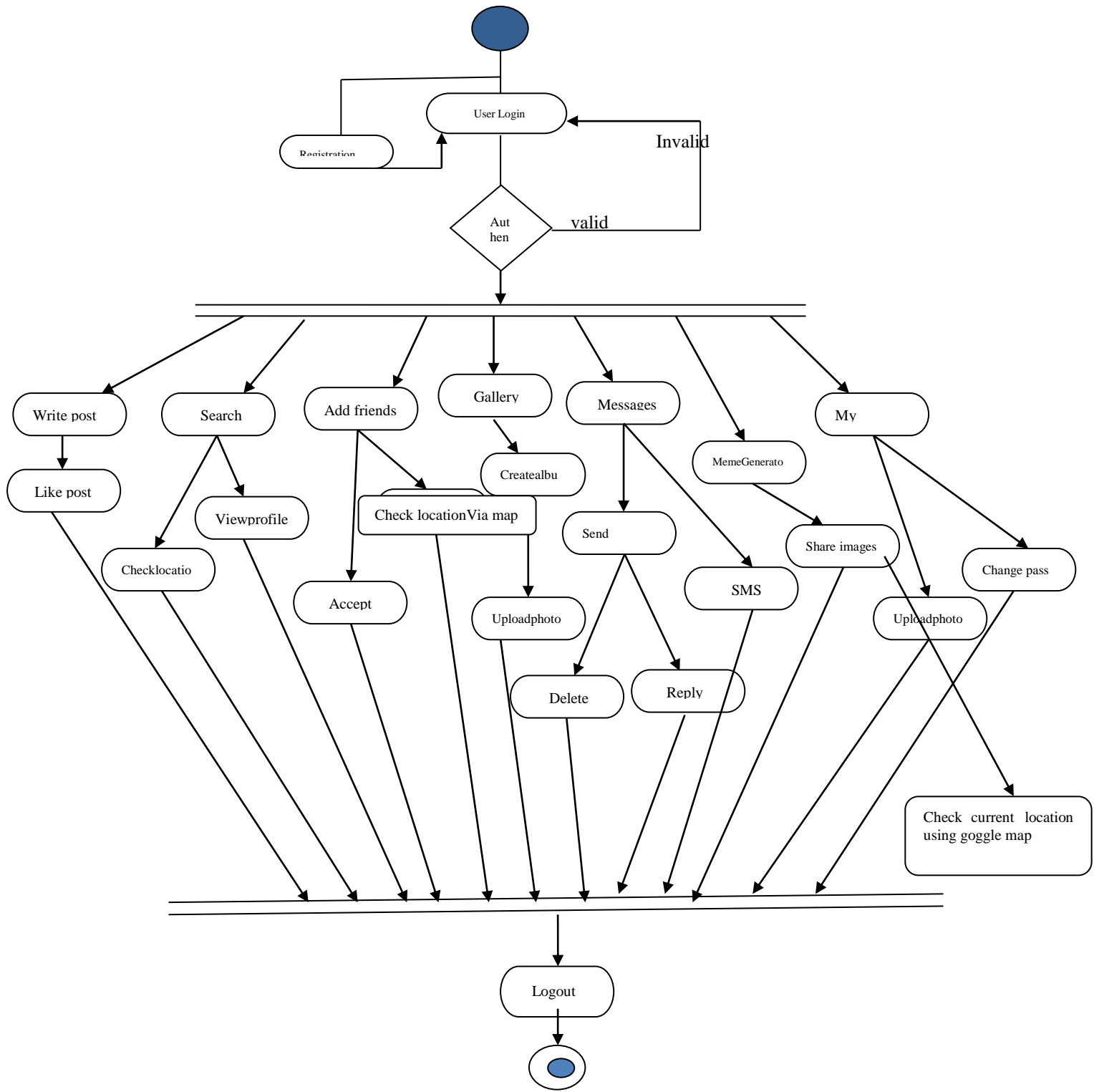

*Fig. 4. The Activity Diagram of the Proposed System*

The fig 4 is the activity diagram which illustrates the Homepage process. The first action occurs when the user makes a post by typing some text or selecting any media file.
Next, the system checks for network connectivity. If the network connectivity is poor, an error message is displayed. If the network is good, the post is uploaded and saved to the MySQL database. Finally, a confirmation message telling the user that the post has been sent is displayed. Other processes involves Adding or deleting informations from his profile, searching and accepting friends, commenting on posts by friends and chatting with friends and mostly of checking the current location of the user using goggle map.
Fig 5 is the use case diagram of the proposed system. The actor which defines the user starts the application on a web browser. A sign up form is displayed. Next, if the user is not registered, the user clicks on the signup button, which is the Login page where he inputs his email and password.
The user's input is processed and checks for validation. If the input is not valid, the user is taken back to the login page and if the input is valid, then finally, the Home page is displayed, where the user can have access to create post, receive a post, update a post and manage account etc.while the user's data are stored in the MySQL database.
The system is introduced when the user opens his or her friend's profile, a button to check the current location of the user is designed, he clicks the button and the current country, state or city is been displayed via Google map.

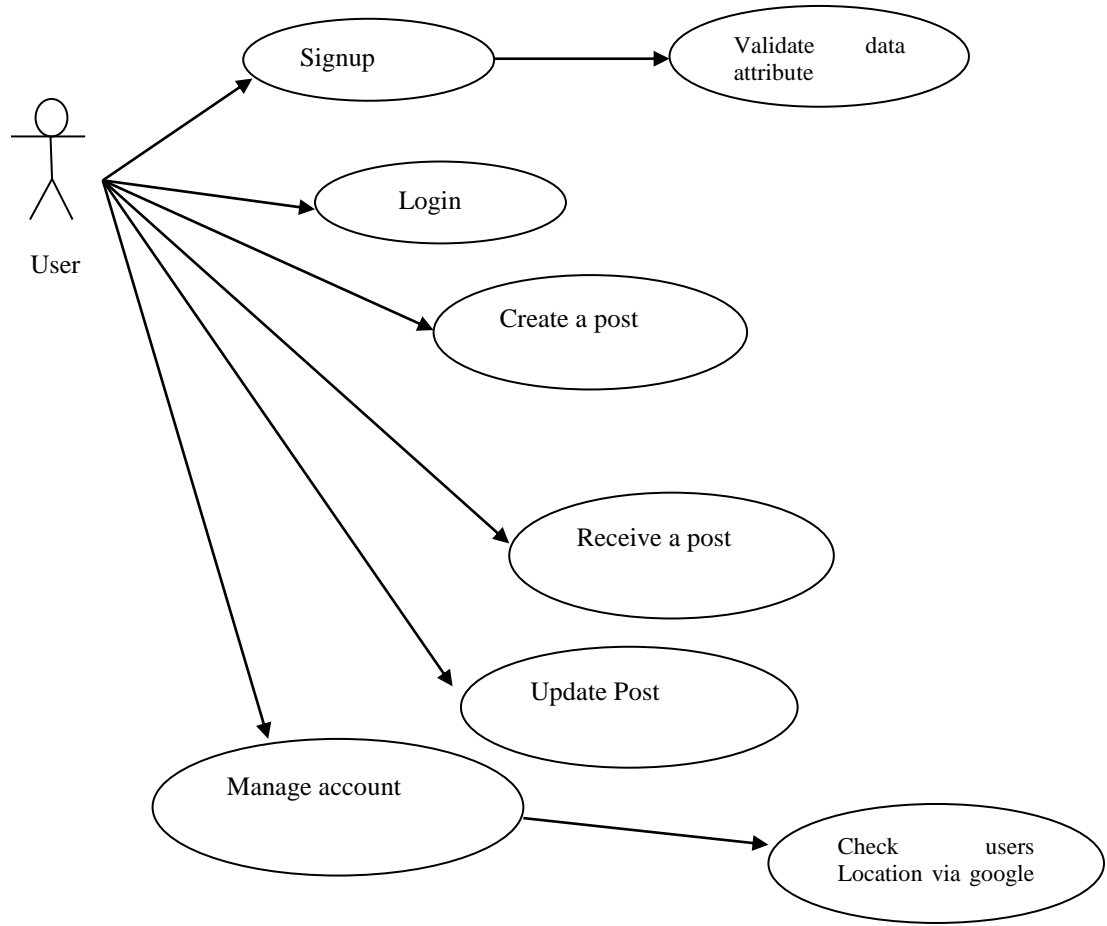

*Fig. 5. Use Case diagram of the Proposed System*

## 4. EXPERIMENT AND RESULTS

This involves the practical method of putting into work all the theoretical design of the proposed model. The enhanced geo-location model for social network communication system was designed using an Object-Oriented software development methodology and implemented using thewith MySQL. A lot of modules were created in the design and implementation of the proposed geo-location social media network system and the results achieved in several phases are displayed in figures 6, 7, 8, 9 and 10.

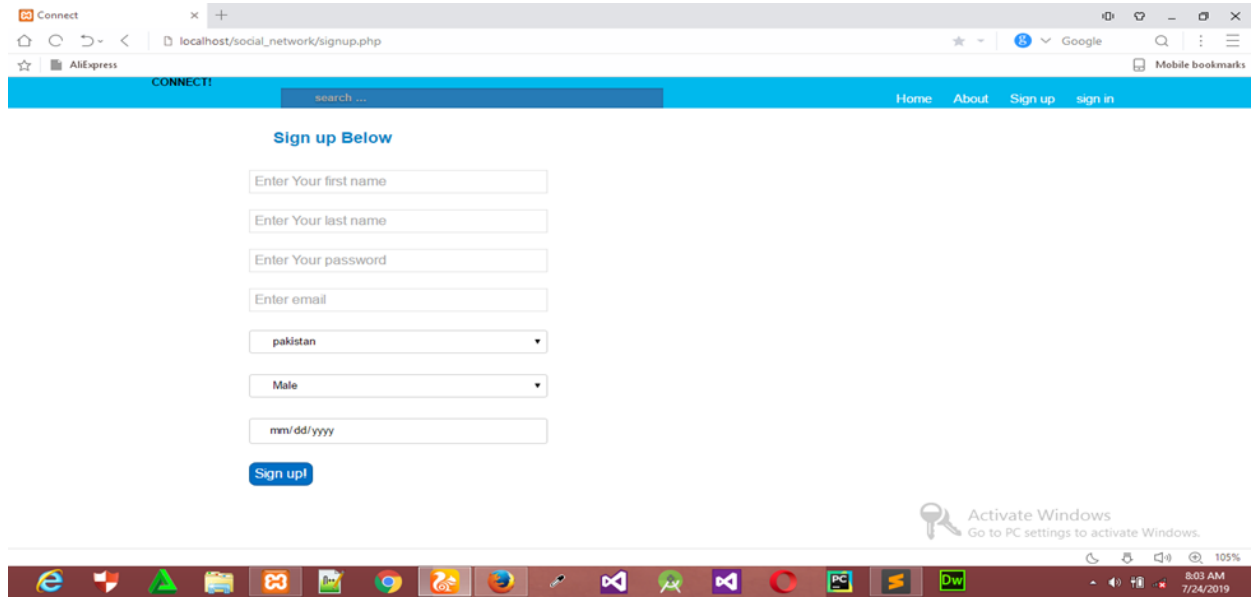

*Fig. 6. Proposed System Signup page*

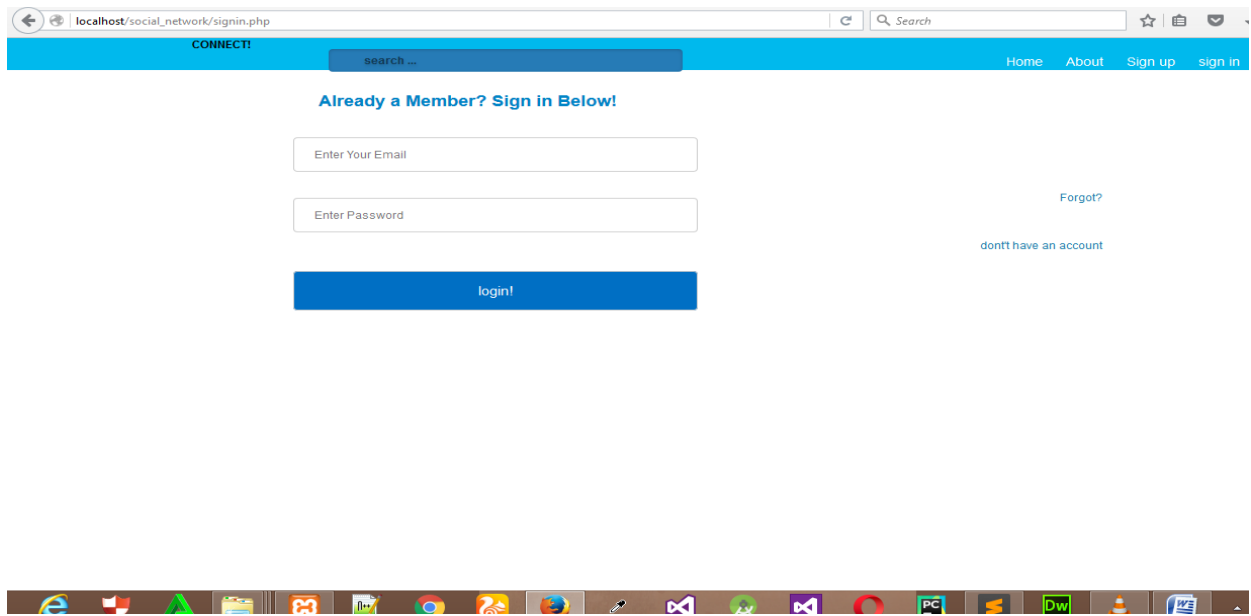

*Fig. 7. Proposed System Login page*

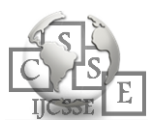

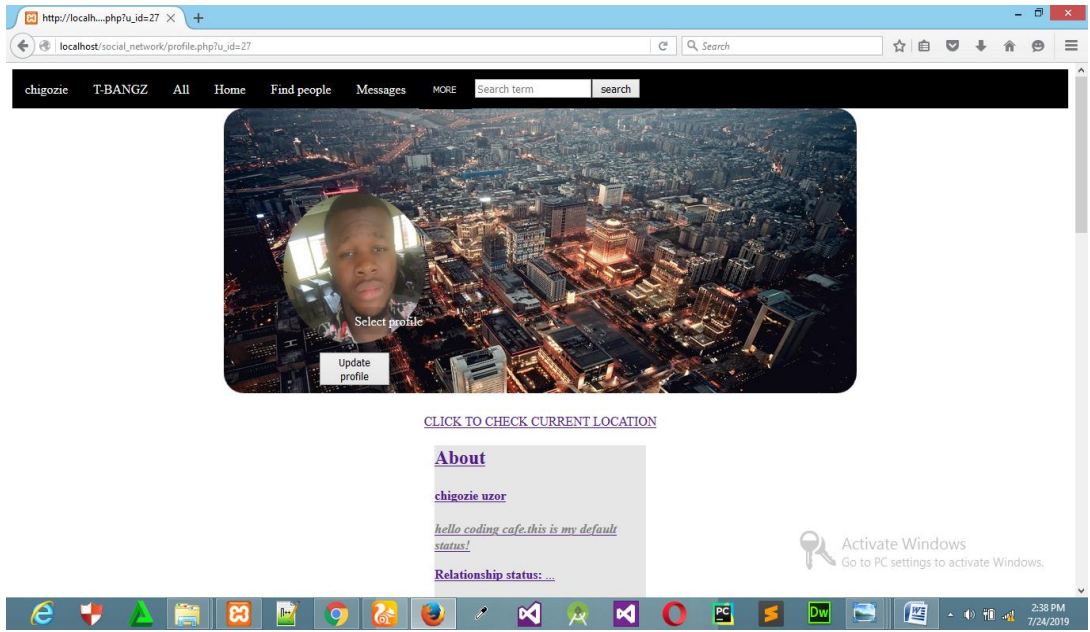

*Fig. 8. System Information Search Module*

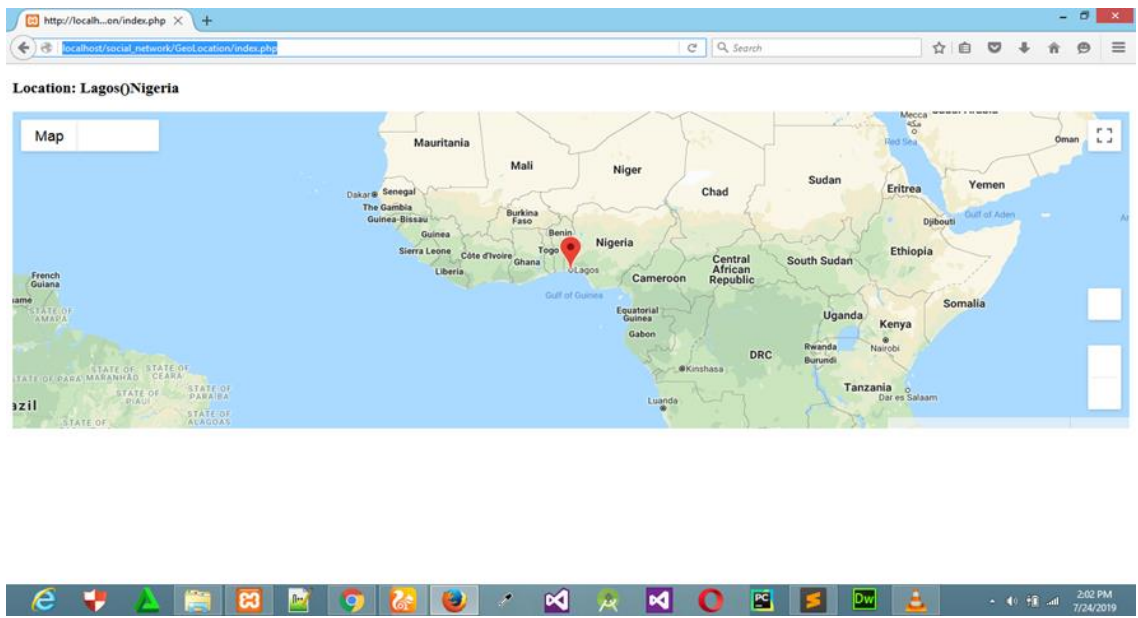

*Fig. 9. System Geo-location Module*

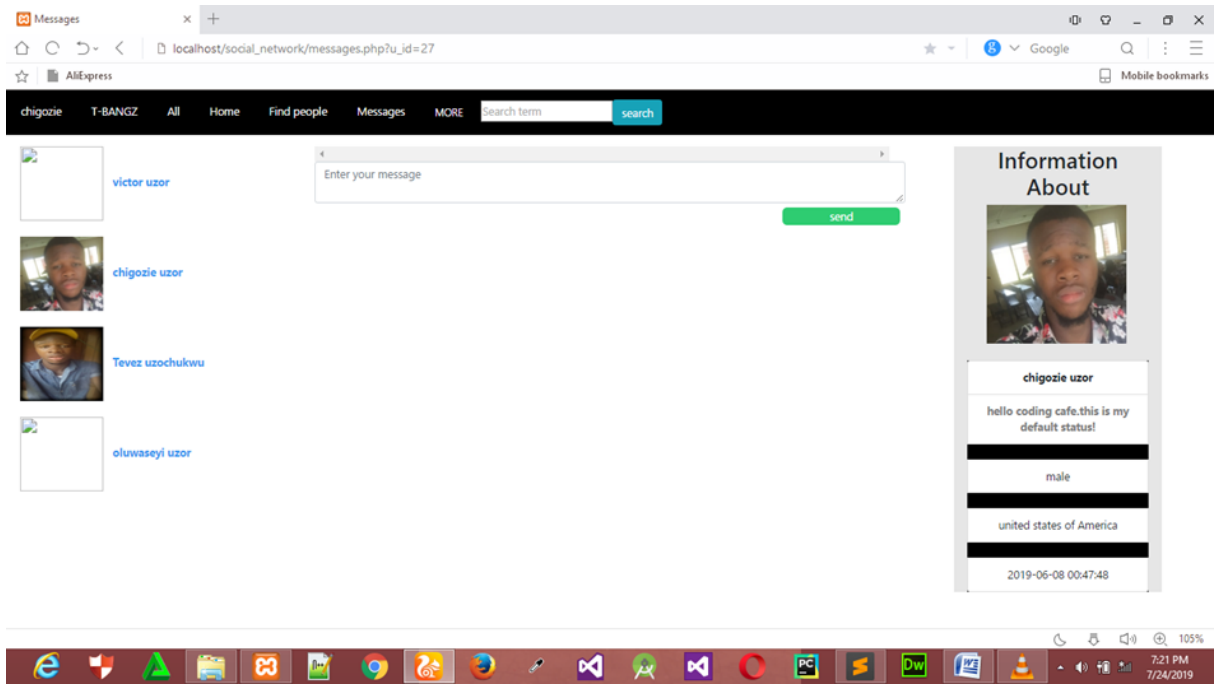

*Fig. 10. Displays Various Users In The Network*

## 5. DISCUSSION OF RESULTS

Fig 6 illustrates the signup page. In this page the user is expected to fill in the form displayed on the screen which consist of the user's data such as firstname, lastname, password, email address, country, gender and date of birth. After that, he clicks on the sign up button which posts the information in the database then it displays a welldone, you are good to go message, which after then, the user goes to the sign up page.

Fig 7 is the login page. This page comes after the user is done with the sign up page. A form is displayed which requests for the user's email address and password. The email address field must contain a valid email address which is registered in the sign up page and the password field does not accept any password that is less than 9 values. After the user is done with filling the details, he clicks on the login button which he is later redirected to the home page if the information is correct; else a form is displayed saying wrong email address or password.

Fig 8 displays the overall information of the user which includes; the personal details and the post by the user. As shown in fig 8, the user's data can be updated. The button labelled "click to check current location" directs a friend or user to a page which displays the current location of the user which entails the country and city. This is important as it reduces the rate of fraud, crime and criminality by fully knowing the location of the user as depicted in fig 9.

Fig 10 displays various users in the network. The user on the platform selects the friends he wants to chat with and starts up a conversation, he sends a message to his friend, in turn the friend when in an online mode sees the message and replies if so desires. Multiple friends are allowed to participate in this social network platform and their locations displayed at the same time instantly.

## 6. CONCLUSION

In this work, an advanced and secured geo-location model for social network communication system has been designed and implemented. The system was designed using an Object-Oriented software development methodology and implemented using the server-based scripting language - PHP, Cascading Style Sheets (CSS) and backend with MySQL. An effective and normalized database is developed for the storage of user's information.

The proposed system has been tested and proved effective. It tracks geographical (latitudinal and longitudinal) location of any user in the system. The system when adopted will certainly help the government and security agencies fight recent security challenges in the country.

## ACKNOWLEDGEMENTS

The authors would like to appreciate the reviewers of this paper, for their useful comments and contributions which added more to the quality of the work.

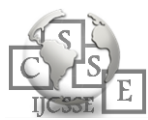